\documentclass[superscriptaddress,twocolumn,amsmath,amssymb]{revtex4-2}
\usepackage{graphicx}
\usepackage{dcolumn}
\usepackage{hyperref}
\usepackage{amsmath}
\usepackage{amssymb}
\usepackage{amsfonts}
\usepackage{mathtools}
\usepackage{color}
\usepackage{datetime}
\usepackage{footnote}
\usepackage[normalem]{ulem}
\usepackage{dcolumn}
 \usepackage{ragged2e}
 \usepackage{sidecap}
 \usepackage{setspace}
 \usepackage{multirow}
\usepackage{xcolor}
 \usepackage{flushend}
\usepackage{cleveref}
\usepackage{microtype}
\usepackage{siunitx}
\usepackage{algorithm}
\usepackage{algpseudocode}
\usepackage[dvipsnames]{xcolor}

\usepackage{etoolbox}   

\makeatletter
\long\def\@makecaption#1#2{%
  \vskip\abovecaptionskip
  \begingroup
    \small
    \justifying 
    \setlength{\parindent}{0pt}%
    \setlength{\leftskip}{0pt}%
    \setlength{\rightskip}{0pt}%
    \textbf{#1.}~#2\par
  \endgroup
  \vskip\belowcaptionskip
}
\makeatother

\usepackage{braket}
\UseRawInputEncoding
\usepackage[T1]{fontenc} 
\usepackage[utf8]{inputenc}

\DeclareMathOperator{\tr}{tr}

\begin{abstract}
Quantum sensors offer significant advantages over classical devices in spatial resolution and sensitivity, enabling transformative applications across materials science, healthcare, and beyond. Their practical performance, however, is often constrained by unmodelled effects, including noise, imperfect state preparation, and non-ideal control fields. In this work, we report the first experimental implementation of a graybox modelling strategy for a solid-state open quantum system. The graybox framework integrates a physics-based system model with a data-driven description of experimental imperfections, achieving higher fidelity than purely analytical (whitebox) approaches while requiring fewer training resources than fully deep-learning blackbox models. We experimentally validate the method on the task of estimating a static magnetic field using a single‑spin quantum sensor, performing Bayesian inference with a graybox model trained on prior experimental data. Using roughly 10,000 training datapoints, the graybox model yields several orders of magnitude improvement in mean squared error over the corresponding physics-only model, as well as significant improvement over a blackbox model with comparable size. These results are broadly applicable to a wide range of quantum sensing platforms, not limited to single-spin systems, and are particularly valuable for real-time adaptive protocols, where model inaccuracies can otherwise lead to suboptimal control and degraded performance.

\end{abstract}

\begin{document}
\title{Bayesian quantum sensing using graybox machine learning}

\author{Akram Youssry}
\address{Quantum Photonics Laboratory and Centre for Quantum Computation and Communication Technology, RMIT University, Melbourne, VIC 3000, Australia}
\affiliation{School of Electrical Engineering and Telecommunications, UNSW Sydney, Sydney, NSW 2052, Australia}

\author{Stefan Todd}
\address{
Institute of Photonics and Quantum Sciences, SUPA,
School of Engineering and Physical Sciences, Heriot-Watt University, Edinburgh EH14 4AS, UK}

\author{Patrick Murton}
\address{
Institute of Photonics and Quantum Sciences, SUPA,
School of Engineering and Physical Sciences, Heriot-Watt University, Edinburgh EH14 4AS, UK}

\author{Muhammad Junaid Arshad}
\address{
Institute of Photonics and Quantum Sciences, SUPA,
School of Engineering and Physical Sciences, Heriot-Watt University, Edinburgh EH14 4AS, UK}

\author{Nicholas Werren}
\address{
Institute of Photonics and Quantum Sciences, SUPA,
School of Engineering and Physical Sciences, Heriot-Watt University, Edinburgh EH14 4AS, UK}

\author{Alberto Peruzzo}
\email{alberto.peruzzo@gmail.com}
\address{Quantum Photonics Laboratory and Centre for Quantum Computation and Communication Technology, RMIT University, Melbourne, VIC 3000, Australia}
\address{Quandela, Massy, France}

\author{Cristian Bonato}
\email{c.bonato@hw.ac.uk}
\address{
Institute of Photonics and Quantum Sciences, SUPA, School of Engineering and Physical Sciences, Heriot-Watt University, Edinburgh EH14 4AS, UK}

\maketitle

\section{Introduction}

Quantum sensors have emerged as powerful tools that harness quantum effects and/or atomic-scale size to deliver high sensitivity in the measurement of different physical quantities~\cite{degen_QuantumSensing_2017, pezze_AdvancesMultiparameterQuantum_2025}. After demonstrating excellent performance in well-controlled laboratories, quantum sensors are now increasingly deployed in real-world applications ranging from navigation~\cite{muradoglu_QuantumassuredMagneticNavigation_2025} and geophysics~\cite{peters_MeasurementGravitationalAcceleration_1999} to medical imaging~\cite{xia_MagnetoencephalographyAtomicMagnetometer_2006, boto_MovingMagnetoencephalographyRealworld_2018, aslam_QuantumSensorsBiomedical_2023} and fundamental physics~\cite{budker_QuantumSensorsHigh_2022, demille_QuantumSensingMetrology_2024}. 

The operation of quantum sensors entails a parameter estimation problem, where an unknown quantity of interest, such as the direction or the strength of a magnetic field, is indirectly estimated from quantum measurements. In this context, Bayesian inference is becoming increasingly important as a probabilistic framework for evaluating parameters associated to quantum states or quantum dynamics~\cite{granade_RobustOnlineHamiltonian_2012, macieszczak_BayesianQuantumFrequency_2014, granade_PracticalBayesianTomography_2016, granade_PracticalAdaptiveQuantum_2017, ferrie2018bayesian, gentile_LearningModelsQuantum_2021, gebhart_LearningQuantumSystems_2023a, wallace_LearningDynamicsMarkovian_2025, fioroni_LearningAgentbasedApproach_2025, belliardo_MultidimensionalQuantumEstimation_2025}. In quantum systems, where uncertainty and probabilistic outcomes are inherent, Bayesian methods offer a natural way to update beliefs about unknown parameters as new data (measurements) are obtained ~\cite{jones_PrinciplesQuantumInference_1991, buzek_ReconstructionQuantumStates_1998}. Further, a Bayesian approach allows for optimal parameter estimation and control even in the presence of noise and limited data ~\cite{blume-kohout_OptimalReliableEstimation_2010, christandl_ReliableQuantumState_2012, sauvage_OptimalQuantumControl_2020a}. By incorporating prior knowledge or assumptions about the system, Bayesian estimation involves updating a posterior probability distribution of the parameter of interest given the observations obtained so far. The procedure requires the use of a ``likelihood'' function that quantifies how well the model of the device is in relation to the observed data.  The posterior probability distribution can be used to compute optimal settings for the next measurements~\cite{bonato_optimized_2015, gebhart_LearningQuantumSystems_2023a}, through Bayesian experiment design~\cite{rainforth_ModernBayesianExperimental_2024}. This can be achieved by maximizing the information gain~\cite{dushenko_SequentialBayesianExperiment_2020, varona-uriarte_ReducingSensingTime_2026}, or other information-theoretic quantities such as Fisher information~\cite{arshad_RealtimeAdaptiveEstimation_2024b} or through reinforcement learning procedures~\cite{belliardo_model-aware_2024, belliardo_applications_2024}. 

Transitioning from controlled environments to real-world applications of quantum sensors presents significant challenges. Crucially, the need to move beyond idealized, analytical models and account for noise, experimental imperfections, and the dependence of their operating parameters on external quantities such as temperature, pressure, magnetic fields and mechanical vibrations. These non-idealities are especially critical for the success of Bayesian inference and in adaptive sensing schemes~\cite{bonato_optimized_2015, joas_OnlineAdaptiveQuantum_2021,arshad_RealtimeAdaptiveEstimation_2024b, berritta_RealtimeTwoaxisControl_2024, berritta_EfficientQubitCalibration_2025}, where real-time feedback and control decisions rely on the accuracy of the underlying model. In such cases, even small mismatches between the model and the actual sensor behavior can lead to degraded performance or instabilities. 

Developing accurate models that capture these real-world complexities is inherently difficult: noise sources are often poorly characterized, imperfections are hard to control, and the dependence of system parameters on environmental conditions are hard to parametrize. A purely data-driven approach, such as training a neural network to model the entire system, is theoretically possible but often impractical. Such models are expensive to train, require large datasets, and lack interpretability, making them unsuitable for applications where performance, reliability, and physical insight are critical.

An alternative is to use a ``graybox'' (GB) modeling approach~\cite{bohlin_InteractiveSystemIdentification_1991, bohlin_PracticalGreyboxProcess_2006}, that combines the strengths of both physically-derived ``whitebox'' (WB) and data-driven ``blackbox''(BB) techniques. By embedding known physical laws (WB) into a flexible model structure that comprises data-driven components (BB) to capture unknown noise and imperfections, we construct a hybrid GB framework that is, at the same time, accurate and computationally efficient. This approach has been first proposed in quantum technology for characterizing photonic quantum circuits~\cite{youssry_ModelingControlReconfigurable_2020, youssry_ExperimentalGrayboxQuantum_2024}, the identification and control of open quantum systems~\cite{youssry_CharacterizationControlOpen_2020,youssry_MultiaxisControlQubit_2023, auza_QuantumControlPresence_2024, mayevsky_QuantumEngineeringQudits_2026, sareen_SingularityfreeDynamicalInvariantsbased_2026, pathumsoot2025probabilistic, pathumsoot2025graybox}, noise detection using spectator qubits~\cite{youssry_NoiseDetectionSpectator_2023}, and geometric gate synthesis~\cite{perrier_QuantumGeometricMachine_2020}.

In this paper, we address the problem of quantum sensing in realistic environments by proposing a framework that integrates GB modelling with Bayesian parameter estimation. After introducing the quantum sensing problem in Section \ref{sec:background}, in Section \ref{sec:WB} we show how to build a rigorous mathematical model for the sensor dynamics that can capture practical non-idealities and imperfections using the noise operator formalism~\cite{youssry_CharacterizationControlOpen_2020}. Next, in Section \ref{sec:GB}, we show how to build and train a GB model that can efficiently model the sensor, and how to integrate it with the Bayesian estimation procedure. The proposed architecture is customized to the sensing application providing a simplified design compared to~\cite{youssry_CharacterizationControlOpen_2020, youssry_MultiaxisControlQubit_2023, youssry_NoiseDetectionSpectator_2023, auza_QuantumControlPresence_2024, mayevsky_QuantumEngineeringQudits_2026, sareen_SingularityfreeDynamicalInvariantsbased_2026}, with the time-order evolution layers replaced by a simpler Ramsey circuit evolution computation. In Section~\ref{sec:experiments}, we experimentally demonstrate the proposed method on a quantum sensor based on the single electron spin associated with an NV center in diamond, used to measure the strength of an applied static magnetic field. This constitutes the first experimental realization of the GB approach in an open quantum system, in contrast to previous works~\cite{youssry_CharacterizationControlOpen_2020, youssry_MultiaxisControlQubit_2023, youssry_NoiseDetectionSpectator_2023, auza_QuantumControlPresence_2024, mayevsky_QuantumEngineeringQudits_2026, sareen_SingularityfreeDynamicalInvariantsbased_2026} where only numerical simulations were presented. 

Our results show that the GB model outperforms the standard WB models commonly used for NV sensors, in terms of the estimation accuracy. This work establishes a framework to improve the performance of quantum sensors in realistic environments, and could easily be extended to different quantum sensors and/or sensing pulse sequences. Moreover, it provides a key building block for implementing adaptive, real-time feedback control schemes, which critically depend on accurate predictive models. Even modest model inaccuracies can severely degrade control performance, leading to unstable operation or outright failure of the control loop, so that inclusion of noise and imperfection in the model is crucial.

\section{Background}
\label{sec:background}
\subsection{Frequency estimation by a Ramsey experiment}
\label{sec:ramsey}
As an example, we consider the problem of estimating the precession frequency $f_B$ of a single qubit ~\cite{macieszczak_BayesianQuantumFrequency_2014} through a Ramsey experiment. The qubit, prepared in $\ket{0}$, is initialized in an equal superposition $\frac{1}{\sqrt{2}} \left( \ket{0}+\ket{1} \right)$ by a $\pi/2$ pulse. It then evolves under the free evolution Hamiltonian $2 \pi f_B \hat{\sigma}_z$ (we take $\hbar = 1$) for a time $t$, after which a second $\pi/2$ is applied, with phase $\phi$. This second pulse translates the information encoded in the accumulated relative phase between $\ket{0}$ and $\ket{1}$ into a difference in the amplitude of the respective populations. At the end of the sequence, the probability of measurement outcome $d \in \lbrace 0, 1 \rbrace$ is:
\begin{align}
\label{eq:likelihood}
P (d|f_B) = \frac{1}{2} \lbrace 1 - (-1)^d e^{-(t/T_2^*)^2}\cos \left( 2\pi f_B t + \phi \right) \rbrace.
\end{align}
 Here we assume the qubit is subject to static or slowly-varying Gaussian-distributed classical noise, resulting in a Gaussian dephasing term, with dephasing time $T_2^*$. An example of this is the detection of a static magnetic field $B_z$ with a single electron spin. The field $B_z$ results in spin precession around $z$ at the Larmor frequency $f_B = \gamma_e B_z$, where $\gamma_e$ is the electron spin gyromagnetic ratio ($\gamma_e \sim 28$ MHz/mT).

\subsection{Imperfect qubit readout}
In general, the qubit readout can be imperfect as there is a non-zero probability of a detector click when not expected, e.g. due to detector noise. Additionally, losses in the detection system can result in no detector click when expected. These imperfections can be accounted for as ~\cite{dinani_BayesianEstimationQuantum_2019a}:
\begin{align}
    P_{\text{cl}} (f_B) = \pi_0 P (0|f_B) + \pi_{1}P (1|f_B),
    \label{eq:p_cl}
\end{align}
where $\pi_{0/1}$ is a factor that represents the probability of a detector click when the qubit is in $\ket{0}$/$\ket{1}$, respectively. These factors can be experimentally calibrated. 
This equation can also re-expressed as ~\cite{dinani_BayesianEstimationQuantum_2019a}
\begin{align}
    P_{\text{cl}} &= \frac{1}{2}\pi_0(1+\braket{Z}) + \frac{1}{2}\pi_1(1-\braket{Z}) \nonumber \\
    &= \frac{1}{2}(\pi_0+\pi_1)\left(1 + \frac{\pi_0-\pi_1} {\pi_0+\pi_1}\braket{Z}\right) \\
    &:= \alpha(1 + V\braket{Z}) \nonumber
\end{align}
This means that, if we can find a model that can be used to predict $\braket{Z}$, then we can predict $P_{\text{cl}}$. 

In many cases, single-shot projective qubit readout might not be available (i.e. $\pi_{0/1} \ll 1$). This scenario is practically relevant for example in experiments on single spins associated to quantum defects at room temperature. In this case, the experiment is repeated $R$ times, detecting $r$ photons, in what is known as ``averaged'' or ``soft'' readout~\cite{danjou_SoftDecodingQubit_2014, degen_QuantumSensing_2017, dinani_BayesianEstimationQuantum_2019a, zohar_RealtimeFrequencyEstimation_2023a}.

\subsection{Bayesian Estimation}
\label{sec:bayesian}
In the example of frequency estimation presented here, the probability distribution for the frequency $f_B$ is updated using Bayes rule after each binary measurement outcome $d_n$:
\begin{align}
P (f_B|\vec{d}_n) = \frac{P (f_B|\vec{d}_{n-1}) P{(d_n|f_B)}}{\mathcal{N}},
\label{eq:bayes_update}
\end{align}
Here $\vec{d}_n$ is a vector comprising the outcomes of the first $n$ measurements ($\vec{d}_n = \lbrace d_n, d_{n-1} ... d_1 \rbrace$). The term $P{(d_n|f_B)}$, known as the ``likelihood'', is given by Equation \ref{eq:likelihood}. The normalization constant $\mathcal{N}$ is defined as
\begin{align}
    \mathcal{N}=\int_{-\infty}^{\infty}P(f_B|\vec{d}_{n-1})P(d_n|f_B)df_B.
\end{align}
In the soft averaging approach, for a sequence of $n$ batches with all measurements within each batch having the same settings, (i.e., the same $\tau$ and $\phi$), the probability of $f_B$ given the measurement results $\vec{r}_n=(r_1, r_2,..., r_n)$ becomes the quantity of interest. The update equation then becomes
\begin{align}
P (f_B|\vec{r}_n) = \frac{P (f_B|\vec{r}_{n-1}) P{(r_n|f_B)}}{\mathcal{N}'},
\label{eq:bayes_update2}
\end{align}
where
\begin{align}
    \mathcal{N}'=\int_{-\infty}^{\infty}P(f_B|\vec{r}_{n-1})P(r_n|f_B)df_B.
\end{align}
Since the probability of detecting more than one photon in one Ramsey is negligible, the probability of detecting $r$ photons in a batch of $R$ Ramseys can be written as a binomial distribution, i.e.,
\begin{align}
P{\left({r|f_B}\right)}=\left( {\begin{array}{*{20}{c}}
{R}\\
{r}
\end{array}} \right)\left[P_{\textrm{cl}}{\left(f_{B}\right)}\right]^r \left[1-P_{\textrm{cl}}{\left(f_{B}\right)}\right]^{R-r},
\end{align}
where $\left( {\begin{array}{*{20}{c}}
{R}\\
{r}
\end{array}} \right)$ is the binomial coefficient.  In the limit of large $r$ and $R$, the binomial distribution can be approximated as a Gaussian distribution~\cite{dinani_BayesianEstimationQuantum_2019a}:
\begin{align}
    P(r|f_B)\approx\frac{1}{\sqrt{2\pi}\sigma}e^{-\frac{(r-\mu)^2}{2\sigma^2}}
\end{align}
where $\mu= P_{\text{cl}}(f_B) R$, and $\sigma^2=r(R-r)/R$ are the mean and variance of the distribution. 

\begin{figure*}
    \centering
    \includegraphics[width=0.8\linewidth]{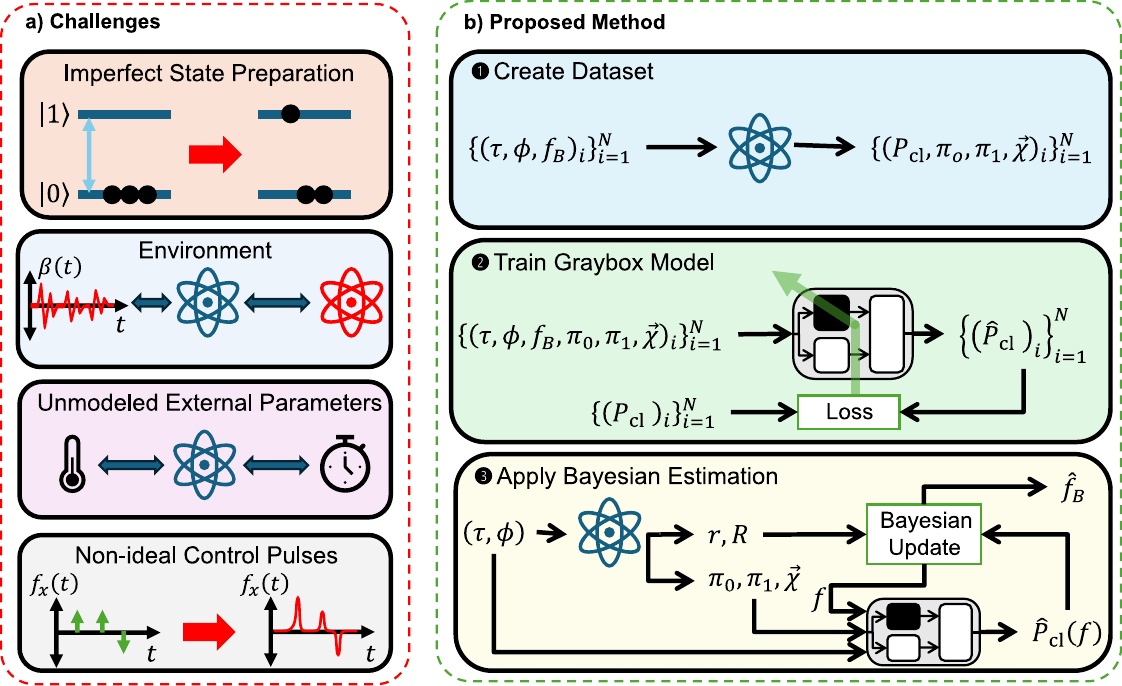}
    \caption{\textbf{Addressing experimental imperfections in quantum sensing.} a) The experimental implementation of quantum sensors faces several non-idealities, including imperfect quantum state initialization; the effect of the (classical and/or quantum) environment, resulting in generally unknown non-Markovian dynamics; unmodeled external factors, such as temperature or vibrations, that can affect the sensor dynamics; and non-ideal control pulses resulting in unknown deviations in the dynamics. b) Our proposed method to target these non-idealities incorporates a graybox (GB) model in a Bayesian estimation procedure. The first step is to create an experimental dataset of Ramsey experiments with randomized parameters (i.e. random pulse parameters $\tau$, $\phi$, $f_B$). The second step is to train a GB model that takes the pulse parameters and calibration coefficients as inputs, and outputs the prediction of the probability of click. The training procedure is based on minimizing a loss function that computes the error between the predictions of the model and the ground truth output. The third step is to apply the Bayesian procedure iteratively to estimate unknown Larmor frequency $\hat{f}_B$ during operation mode. In each iteration, we execute a Ramsey experiment with random delay and phase shift, while the true unknown Larmor frequency $f_B$ is on. The qubit is measured, and the pulse sequence parameters are fed to the trained GB to predict $\hat{P}_{\text{cl}}(f)$ for a set of frequencies $f\in [f_{\min}, f_{\max} ]$ which we assume the true $f_B$ lies in.  This prediction together with the  measured number of clicks $r$ and batch size $R$, are used in the Bayes update rule to find an update posterior distribution of $P(f_B|\vec{r}_n)$, from which the estimate of the Larmor frequency $\hat{f}_B$ is estimated. The iterations are repeated until convergence.}
    \label{fig:protocol}
\end{figure*}

If the likelihood model is not accurate enough,  in other words, $P_{\text{cl}}(d|f_B)$ is not faithfully represented by Equation \ref{eq:likelihood}, the Bayesian procedure will not succeed. 
A failure of the likelihood function to fully model the experiment can be attributed to four main challenges, as depicted in Figure \ref{fig:protocol}a:
\begin{itemize}
    \item \textit{Imperfect initial state preparation}: The initial state of the qubit is a mixed state rather than a pure state.
    \item \textit{Complex (or unknown) interaction with the environment}: In most platforms, qubits interact with their environment resulting in complex dynamics that can be difficult to model exactly.
    \item \textit{Un-modeled experimental settings}: External parameters such as temperature or mechanical vibrations can induce changes in the dynamics that are very hard to measure and model mathematically.
    \item \textit{Non-ideal control}: imperfections in signal generation and delivery (for example in the electronics chain in the case of microwaves) will cause pulse distortions, which are generally challenging to characterize experimentally. 
\end{itemize}

In this paper, we aim to address these challenges by using a machine-learning approach that can capture these imperfections directly from data, without requiring a mathematical model, and providing a more accurate likelihood function $P(r|f_B)$. This will be described next. 

\section{Methods}
\label{sec:methods}

\subsection{``Whitebox'' quantum sensor model}
\label{sec:WB}
In this section, we discuss how one can construct a physical model for the experiment described in Section \ref{sec:ramsey}, including all known sources of errors and imperfections.

A first common imperfection is state-preparation errors. While we target qubit initialization in $\rho(0)=\ket{0}\!\!\bra{0}$, a residue of the ensemble can remain in the state $\ket{1}\!\!\bra{1}$, resulting in the physical initial state 
\begin{align}
    \mathcal{E}(\rho(0))&=(1-\epsilon)\ket{0}\!\!\bra{0} + \epsilon\ket{1}\!\!\bra{1},
\end{align}
where $0<\epsilon\ll1$ is a small positive factor encoding the quality of the procedure (i.e. a perfect preparation corresponds to $\epsilon=0$). 

In Supplementary Note 1, we show how this imperfect initial state could result, for example, from evolution of the state for $t \in[0,T_p]$ under a stochastic Hamiltonian of the form
\begin{align}
    H_{\text{prep}}(t)&=\begin{cases}
        \beta(t)\sigma_x, \quad &0\le t \le T_p\\
        0, \quad &t>T_p
    \end{cases}
\end{align}
for a certain random process $\beta(t)$. The remaining terms in the total Hamiltonian can then be defined so that their support is over $[T_p,T]$, rather than $[0,T]$, enabling us to split the evolution trajectory of $\rho(0)$ into two sub-trajectories. In the first, the state evolves under the effect of $H_{\text{prep}}(t)$ to model state preparation errors. In the second, the system continues to evolve under the remaining Hamiltonian terms representing the actual physical processes of interest (i.e. applying control and interaction with environment). These remaining terms can be written as:
\begin{itemize}
    \item $H_{\text{ctrl}}(t)$ is the system control Hamiltonian, encoding the effects of the static and dynamic magnetic fields, accounting for the preparation of the superposition state, the evolution under the static magnetic field, and the measurement operation. It will take the form
    \begin{align}
        H_{\text{ctrl}}(t) = 2\pi f_B \sigma_z + f_x(t)\sigma_x+ f_y(t)\sigma_y,
    \end{align}
    where $f_x(t)$ and $f_y(t)$ are the functions representing the components along the $X$- and $Y$-axes of the pulses applied to the qubit. Generally, near-perfect pulses are created by an arbitrary waveform generator (AWG) with high temporal and amplitude accuracy. However, the actual waveforms delivered to the qubit are typically distorted by a variety of physical effects, including attenuation and dispersion from propagation in cables, reflections from imperfectly-matched antennas and equipment, the finite bandwidth of the associated electronics (e.g. mixers, amplifiers), etc. In practice, the exact pulse waveforms driving the qubit are generally unknown and not trivial to characterize completely. 
\item $H_B(t)+H_{SB}(t)$ is the Hamiltonian of the quantum bath, and the interaction Hamiltonian between the system and the bath, respectively. These two terms represent quantum noise affecting the system, which for instance can originate from unwanted interactions with nuclear spins associated to the atoms in the crystal lattice, phonons, two-level fluctuators, etc.

\item $H_{\text{stoch}}(t)$ is a stochastic system operator encoding classical noise affecting the system. This can originate from unwanted electromagnetic interference, or noise from the electronics. 
\item $H_{\text{ext}}(t;\vec{\chi})$: is a general system and/or bath term that represents any changes to any of the other Hamiltonian terms due to some external measurable parameters $\vec{\chi}$ (such as temperature). Here we assume generally that this term is unknown. 
\end{itemize}
The inclusion of those Hamiltonian terms allows modeling Markovian or Non-Markovian dynamics depending on the time-correlation properties of the noise. For instance, if the classical noise in $H_{\text{stoch}}(t)$ is colored, then the resulting dynamics is non-Markovian ~\cite{chenu_QuantumSimulationGeneric_2017}. The total Hamiltonian of the system can then be expressed as 
\begin{align}
    H(t) = H_{\text{Ramsey}}(t;\tau,\phi,f_B) + H_{\text{noise}}(t),
\end{align}
where $H_{\text{Ramsey}}(t; \tau,\phi,f_B)$ is the ideal Hamiltonian corresponding to an ideal Ramsey experiment pulse sequence with a drift of $2\pi f_B$, time delay $\tau$ between the pulses, and a phase shift $\phi$. On the other hand, the term
\begin{align}
    H_{\text{noise}}(t) &= \left(H_{\text{ctrl}}(t) - H_{\text{Ramsey}}(t)\right) + H_{\text{prep}}(t)\nonumber \\
    &+H_B(t) + H_{SB}(t) + H_{\text{stoch}}(t) + H_{\text{ext}}(t;\vec{\chi})
\end{align}
encodes all the imperfections including state preparation errors, noisy interaction with classical and quantum environments, effects due to external parameters, and non-ideal pulse effects (such as finite-width pulses instead of impulses; and pulse distortions). Note the first term here is the difference between the ideal Ramsey pulse sequence and the actual pulse sequence applied. In general, these terms are unknown, and can be difficult to characterize individually. 

Having this split of the total Hamiltonian, we can then apply the noise operator formalism~\cite{youssry_CharacterizationControlOpen_2020}, to obtain a compact representation of the expectation of any system observable $O$,
\begin{align}
    \braket{O(T)}=\tr\left(V_O(T;\tau, \phi, f_B, \vec{\chi}) \tilde{\rho}(T) O\right),
    \label{equ:obs}
\end{align}
where
\begin{align}
    \tilde{\rho}(T)=U_{\text{Ramsey}}(\tau,\phi, f_B)\rho(0)U_{\text{Ramsey}}^{\dagger}(\tau,\phi,f_B)
\end{align}
is the ideal (closed-system) state evolving under the Ramsey experiment, and $V_O(T;\tau,\phi,f_B, \vec{\chi})$ is a system operator that encodes all the noise effects and will depend on the ideal pulse sequence parameters $\tau$, $\phi$, and $f_B$. We assume that the system and bath initial joint state are separable, i.e. $\rho_{SB}(0)=\rho(0)\otimes\rho_B$.  The ideal Ramsey experiment unitary 
$U_{\text{Ramsey}}(\tau,\phi,f_B)$ can be expressed as 
\begin{align}
    U_{\text{Ramsey}}(\tau,\phi,f_B) = R_X(\pi/2)R_Z(\theta)R_X(\pi/2),
\end{align}
where $R_X$ and $R_Z$ are the rotation about X-axis and Z-axis, and 
\begin{align} 
    \theta = 2\pi f_B\tau + \phi.
\end{align}
The noise operator is formally defined as 
\begin{align}
    V_O(T) = O^{-1}\left\langle\tr_B\left(\tilde{U}_I^{\dagger}(T) O \tilde{U}_I(T) \rho_B\right)\right\rangle_c,
    \label{equ:Vodef}
\end{align}
where 
\begin{align}
    \tilde{U}_I(T):=U_I(T)U_{\text{Ramsey}}(\tau,\phi,f_B),
\end{align}
and
\begin{align}
    U_I(T)&:=\mathcal{T}e^{-i\int_0^T{H_I(s)ds}}
\end{align} 
is the time-ordered exponential of the interaction Hamiltonian $H_I(t)=U_{\text{Ramsey}}^{\dagger}(\tau,\phi,f_B)H_{\text{noise}}(t)U_{\text{Ramsey}}(\tau,\phi,f_B)$ in the interaction picture, and $\braket{\cdot}_c$ is a classical average over all classical noise processes in $H(t)$. We can see that when the noise operator is the system identity operator, $V_O(T)=\mathbb{I}_s$, then we retrieve back the closed-system ideal dynamics under the Ramsey experiment. 

While the noise operator provides an elegant representation of deviation from ideal dynamics, it is generally hard or intractable to compute analytically in a closed-form. We have to resort to either imposing strong assumptions on the noise, or use perturbative expansions~\cite{sareen_SingularityfreeDynamicalInvariantsbased_2026}. However, this will also fail if the noise is strongly-coupled to the system, or the evolution time is long~\cite{auza_QuantumControlPresence_2024}. Moreover, the computations require knowing all the noise terms of $H_{\text{noise}}(t)$, which is challenging to know a priori. Thus, in practical settings we need to estimate the $V_O(T)$ operator indirectly from experimental measurements by fitting a model. Note that $V_O(T)$ itself is in general non-Hermitian, and cannot be measured directly in experiments. 

\subsection{Graybox Model}
\label{sec:GB}
To address the challenges associated with the ``whitebox'' model derived exclusively from physics principles described by Equation \ref{equ:Vodef} is challenging, as explained, we deploy here a ``Graybox'' (GB) approach that combines a physics-based model with ``blackbox'' description of noise and imperfection. Compared to previous work from some of our co-authors, such as~\cite{youssry_CharacterizationControlOpen_2020, youssry_MultiaxisControlQubit_2023, auza_QuantumControlPresence_2024}, where we modeled only imperfections due to external environment using the framework of noise operators, here we rigorously integrate other imperfections, as described in Section \ref{sec:WB}.

\subsubsection{Constructing the GB model}

\begin{figure}
    \centering
    \includegraphics[width=\linewidth]{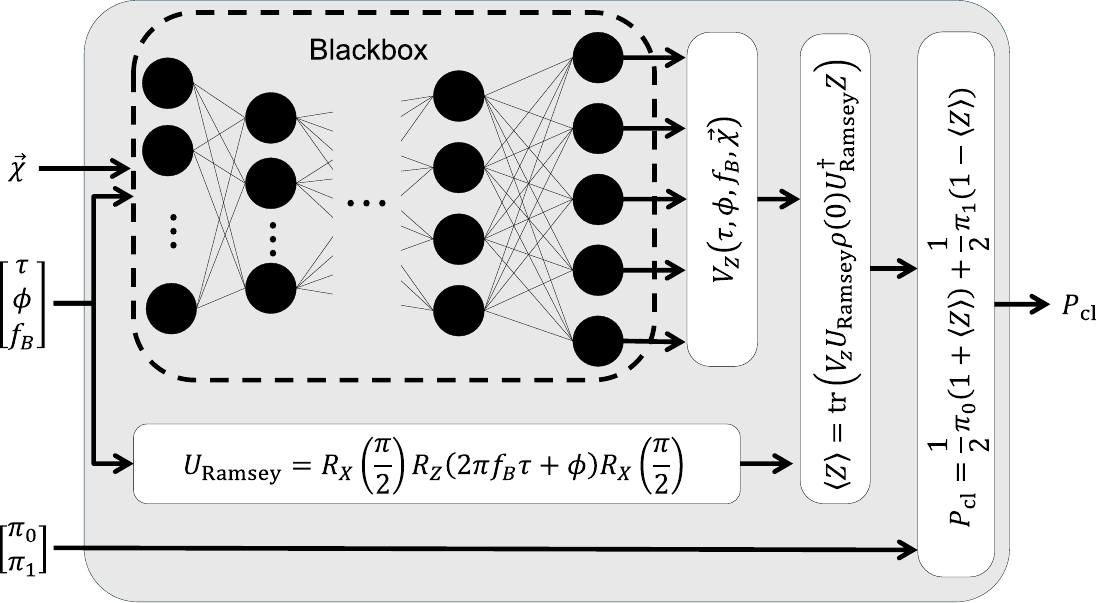}
    \caption{\textbf{The proposed (GB) model architecture.} The input to the model is the ideal Ramsey pulse sequence parameters: (time delay $\tau$, phase shift $\phi$, and Larmor frequency $f_B$, the external parameters $\vec{\chi}$, as well as the readout calibration coefficients $\pi_0$ and $\pi_1$. The pulse parameters and the external parameters are processed through a BB in the form of a neural network of 9 layers. The output of the BB is the matrix parameterization of the noise operator $V_Z$, which is then used to reconstruct the operator itself in the next WB layer. The pulse parameters are also processed in a WB layer that computes the ideal Ramsey circuit unitary $U_{\text{Ramsey}}$. The two paths then merge in WB layer that computes the Pauli-Z expectation $\braket{Z}$. The output layer of the model then computes the probability of click $P_{\text{cl}}$ given the calibration coefficients and the expectation value. }
    \label{fig:GB}
\end{figure}

We define a GB model having as inputs the ideal pulse sequence parameters $(\tau,\phi,f_B)$, a vector of parameters $\vec{\chi}$, and the detector calibration coefficients $(\pi_0,\pi_1)$ (defined in Equation \ref{eq:p_cl}), as output the prediction of the probability of getting a click $P_{\text{cl}}$. We choose the structure of the GB model, sketched in Figure \ref{fig:GB},  to reflect the computation in Equations \ref{eq:p_cl} and \ref{equ:obs}. 

We first have a BB block, in which the pulse sequence and external parameters are processed through a neural network consisting of 8 layers of 1024, 512, 128, 64, 32, 16, 8, 4 nodes, respectively, and a hyperbolic tangent activation. The output of this network is then connected to another neural layer consisting of 3 nodes of linear activation and 2 nodes of hyperbolic tangent activation. The output of these nodes represents the matrix parameterization of the noise operator estimate $\hat{V}_Z(T)$. In other words, the structure described so far maps the pulse sequence parameters to the matrix parameterization of the noise operator estimate. 

We then add WB layers to the BB block, to complete the GB structure. First, we have a layer that reconstructs the $\hat{V}_{Z}(T)$ as a matrix, given the output of the BB part. Next, we add a layer that reconstructs the ideal unitary operator for the Ramsey experiment given the pulse parameters, i.e. this layer is directly connected to the GB input. Finally, these two paths are merged into the final layer, which computes $P_{\text{cl}}$ by first computing $\braket{Z}$ using Equation \ref{equ:obs} and then calculating $P_{cl}$ with Equation \ref{eq:p_cl}, using the values for $\lbrace \pi_0, \pi_1 \rbrace$. 

This structure assumes  that the effect of the external experimental parameters $\vec{\chi}$ on the qubit dynamics is unknown, and thus connects directly to the BB block. On the other hand, if the dependence is known, then the corresponding mathematical relations could be added to the WB part instead of the BB block. 

\subsubsection{Training the GB model}
Once the model is built, the BB block needs to be trained using experimental data. We construct a training dataset consisting of tuples of the form $(\tau,f_B,\phi,\pi_0,\pi_1,\vec{\chi},P_{cl})$, where the calibration coefficients and probability of clicks are measured experimentally. The training examples are used to optimize a loss function of the model: here, we use the logarithm of the MSE between the prediction of the GB $\hat{P}_{\text{cl}}$, and the actual measured probability of click $P_{\text{cl}}$. The loss function is computed over all training examples, and the average is then used to update the weights of the neural networks inside the GB using backpropagation, through the  Adam optimization algorithm ~\cite{kingma_AdamMethodStochastic_2014}. Finally, we check the generalization performance of the trained model by evaluating the same loss function over the testing examples that have not been included in the training procedure. In general, a desired property is to have the values of the training and testing losses to be close, and both decreasing with the number of optimization iterations. However, the ultimate test for the GB model is when it is utilized for a given task. For example in~\cite{auza_QuantumControlPresence_2024}, the GB model had high MSE for the cases of ultra-strong noise, but still performed well when used for pulse engineering. In this paper, since we focus on sensing, the performance test is the sensing precision when using the GB.

\subsection{Frequency Estimation}
Having the GB model trained on the dataset, it can then be used to predict the likelihood function $P(r|f_B)$ for any frequency $f_B\in[f_{\min},f_{\max}]$, where $f_{\max}$ and $f_{\min}$ are the maximum and minimum frequencies that are  used in the experiment. The Bayesian estimation procedure can then be applied to find an estimate $\hat{f}_B$ of the true frequency given the measurement results. The details of the procedure can be described as follows. We consider two types of priors: a uniform distribution,
\begin{align}
    P_u (f_B) = \begin{cases} \frac{1}{f_{\max}-f_{\min}},& f_B \in [f_{\min},f_{\max}] \\ 0,& \text{otherwise}
    \end{cases},
\end{align}
and a Gaussian prior,
\begin{equation}
    P_g (f_b) = \frac{1}{\sqrt{2\pi s}} e^{-(f_b-m)^2/2s^2}, 
\end{equation}
with $m = \frac{1}{2}\left(f_{\min}+f_{\max}\right)$ and $s = \frac{1}{6}\left(f_{\max} - f_{\min}\right)$. 

After each new measurement, we compute numerically the value of $\mathcal{N}'$, and if it is non-zero, we update the posterior as in Equation \ref{eq:bayes_update2}. Otherwise, we do not update the posterior, and keep it equal to the posterior from the previous iteration. To obtain the point-estimate of the frequency, we choose to use the mean of the posterior distribution, 
\begin{align}
    \hat{f}_B=\int_{-\infty}^{\infty}f_B P(f_B|\vec{r}_n)df_B.
\end{align}
This can be computed at each iteration. As performance metrics, the estimation error
\begin{align}
    \text{E}(\hat{f_B},f_{B}):=(\hat{f}_B-f_{B}
)^2\end{align}
and the variance
\begin{align}
    \mathcal{V} = \int_{-\infty}^{\infty}(f_{B}-\hat{f}_B)^2 P(f_B|\vec{r}_n)df_B
\end{align}
can be computed at each iteration. Finally, the integrals in the previous expressions need to be computed numerically rather than analytically. Hence, we approximate those integrals by discretizing the interval $[f_{\min},f_{\max}]$ into $M$ subintervals and then apply the trapezoidal rule.
We adopt this direct grid-based evaluation of the posterior, rather than Markov Chain Monte Carlo (MCMC) sampling, because it is naturally suited to future adaptive sensing implementations: the discretized posterior can be updated in closed form after every new measurement via Eq. ~\ref{eq:bayes_update2}, without the re-initialization and re-thermalization that MCMC would require at each step. This choice follows established practice in Bayesian quantum parameter estimation; a broader comparison of other approaches (including MCMC) and a discussion of their suitability for real-time protocols can be found in the work of Belliardo et al~\cite{belliardo_MultidimensionalQuantumEstimation_2026}.

\section{Experimental Results}
\label{sec:experiments}

\paragraph*{Experimental Setup:} We demonstrate the technique described above on a quantum sensor implemented with the electronic spin associated to a single nitrogen-vacancy (NV) center in diamond. The NV center is a point defect consisting of a substitutional nitrogen atom adjacent to a lattice vacancy, forming an electronic $S=1$ spin that can be initialized and read out optically at room temperature. Capitalizing on the high spatial resolution enabled by atomic-scale wavefunctions and the sensitivity given by long coherence even at room temperature, NV centers have been used for nanoscale mapping of magnetic ~\cite{budakian_RoadmapNanoscaleMagnetic_2024}, electric  ~\cite{dolde_ElectricfieldSensingUsing_2011a, qiu_NanoscaleElectricField_2022, huxter_ImagingFerroelectricDomains_2023}, strain ~\cite{broadway_MicroscopicImagingStress_2019} and temperature ~\cite{kucsko_NanometrescaleThermometryLiving_2013, toyli_FluorescenceThermometryEnhanced_2013} fields. This has resulted in a large variety of diverse applications, such as probing magnetism and electrical currents in condensed matter systems~\cite{kolkowitz_ProbingJohnsonNoise_2015, ku_ImagingViscousFlow_2020a,  vool_ImagingPhononmediatedHydrodynamic_2021, borst_ObservationControlHybrid_2023,  jayaram_ProbingVortexDynamics_2025}, extending nuclear magnetic resonance to the micro/nanoscales~\cite{staudacher_NuclearMagneticResonance_2013, vandestolpe_Mapping50spinqubitNetwork_2024,  du_SinglemoleculeScaleMagnetic_2024, budakian_RoadmapNanoscaleMagnetic_2024}, monitoring biochemical phenomena in living cells~\cite{lesage_OpticalMagneticImaging_2013, nie_QuantumMonitoringCellular_2021a, mzyk_RelaxometryNitrogenVacancy_2022, shanahan_QBiCBiocompatibleIntegrated_2025} and enhancing the sensitivity of medical diagnostics for early disease detection~\cite{miller_SpinenhancedNanodiamondBiosensing_2020, aslam_QuantumSensorsBiomedical_2023, thomasdecruz_QuantumenhancedNanodiamondRapid_2025}.  

\begin{figure}
    \centering
    \includegraphics[width=\linewidth]{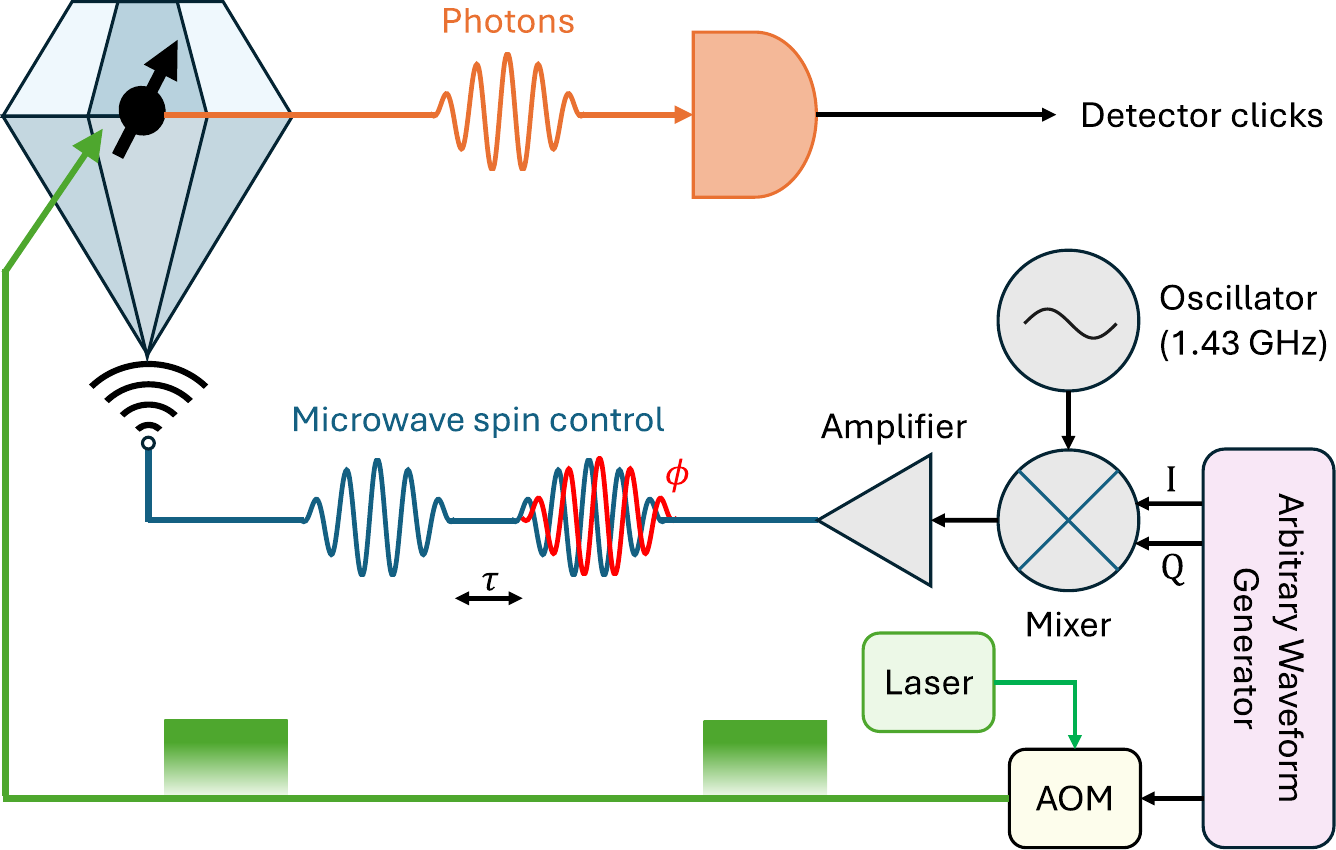}
    \caption{\textbf{Sketch of the NV center experiment.} The electron spin associated to an NV center in diamond is initialised and readout by optical pulses at 532nm, created by an acousto-optic modulator (AOM), and controlled by a microwave pulse sequence. We consider a Ramsey experiment consisting of two $\pi/2$ microwave pulses with tunable inter-pulse delay $\tau$ and relative phase $\phi$ of the second pulse. The microwave pulses are created by single-sideband modulation of a carrier at 1.43 GHz with 40 MHz pulses from an arbitrary waveform generator. All experiments are performed at room temperature. A detailed description of the experimental setup can be found in~\cite{arshad_RealtimeAdaptiveEstimation_2024b}. }
    \label{fig:setting}
\end{figure}

Measurements were carried out on a single NV center in high-purity diamond, at room temperature, using the setup described by MJ Arshad et al~\cite{arshad_RealtimeAdaptiveEstimation_2024b}, sketched on Fig. ~\ref{fig:setting} . We apply a 50 mT magnetic field, along the NV axis, with a permanent SmCo magnet, to polarize the $^{14}$N nuclear spin~\cite{fischer_OpticalPolarizationNuclear_2013}. Both the microscope stage and the permanent magnet are temperature-stabilized (within 10 mK) with a Peltier element driven by a TEC controller with a PID loop (Meerstetter TEC-1091). The NV electron spin is controlled by microwave pulses, created by single-sideband modulation of a 1.43 GHz carrier (R\&S SMBV100A) by 40 MHz pulses generated by a Zurich Instruments HDAWG4 arbitrary waveform generator. The microwave pulses are amplified (Amplifier Research 15S1G6, 15W) and directed to the sample through a coplanar waveguide with a thin copper wire soldered on top of the diamond, to generate the microwave frequency magnetic field required for spin control. The microwave generation and delivery chain is expected to introduce imperfections and non-idealities on the control pulses, through a variety of mechanisms: electronics noise, finite bandwidth of the components, pulse reflections through imperfect impedance matching (particularly the hand-soldered copper wire used as an antenna), etc. These effects are hard to model and characterize, justifying the need for our GB approach.

The experimental data acquisition sequence, fully automated, includes frequent calibration measurements, to monitor closely different parameters used by the GB model. The calibration sequence includes the following steps: \textit{(i)} We first optimize the NV center position in the confocal microscope: we sweep the position of each axis of the three-axes positioning system, record the detected photo-luminescence, fit it with a Gaussian function and move the positioning stage to the center of the Gaussian; \textit{(ii)} We take a coarse optically-detected electron spin resonance spectrum on the NV, to find the resonance frequency. Every 10 sets of 32 Ramsey data-points, we perform a Rabi measurement to identify the duration of the  $\pi/2$-pulse; \textit{(iii)} We measure the values of $\pi_0$ and $\pi_1$, initializing the spin state in $m_s=0$ with a laser pulse (and a microwave $\pi$-pulse to initialize in $m_s=1$) and detecting the photon count rate over a temporal window of 270 ns (with a 400 ns normalization window); \textit{(iv)} we run a Ramsey measurement to identify any residual detuning of the microwave pulses, and adjust the microwave frequency accordingly to set the detuning to zero.

After the calibration step, we randomly select a detuning $f_B$ (corresponding to the emulation of an applied magnetic field $B = f_B/\gamma_e$) and phase $\phi$, and take 32 Ramsey measurements for different values of $\tau$ and phase $\phi$. We choose an exponentially increasing sequence of $\tau$ values ($\tau_k = \tau_0 \cdot 2^k$), as optimal for high-dynamic range magnetometry, as it removes the aliasing of period $1/\tau_k$ that is created when retrieving a frequency from a measured phase ~\cite{higgins_EntanglementfreeHeisenberglimitedPhase_2007, berry_OptimalStatesAlmost_2000, berry_HowPerformMost_2009, bonato_optimized_2015, zohar_RealtimeFrequencyEstimation_2023a}. To minimize the impact of systematic effects such as mechanical drift, we randomize the acquisition order of the $\lbrace \tau_k \rbrace$, so that for example the longer sensing times are not taken always later than the shorter sensing times. Each Ramsey experiment is repeated $R = 5 \times 10^6$ times to accumulate sufficient statistics.

\paragraph*{BB model:} We next compare the GB performance against a pure BB model, to confirm that the GB requires fewer resources to achieve a certain performance level. We choose the BB architecture to largely match the BB part of the GB architecture, i.e. an 9-layer deep neural network of 1024, 512, 128, 64, 32, 16, 8, 4, and 4 nodes respectively with Rectified Linear (ReLU) activation. We find that using this activation instead of the hyperbolic tangent used in the GB architecture results in an improved performance of the BB. This network is then followed by the output layer which is simply a single neuron with sigmoid activation in order to guarantee the output's range is between 0 and 1, matching the range of $P_{\text{cl}}$. The total number of trainable parameters is then $607 669$ as compared to $605 621$ for the GB architecture.

\begin{figure}
    \centering
    \includegraphics[width=0.85\linewidth]{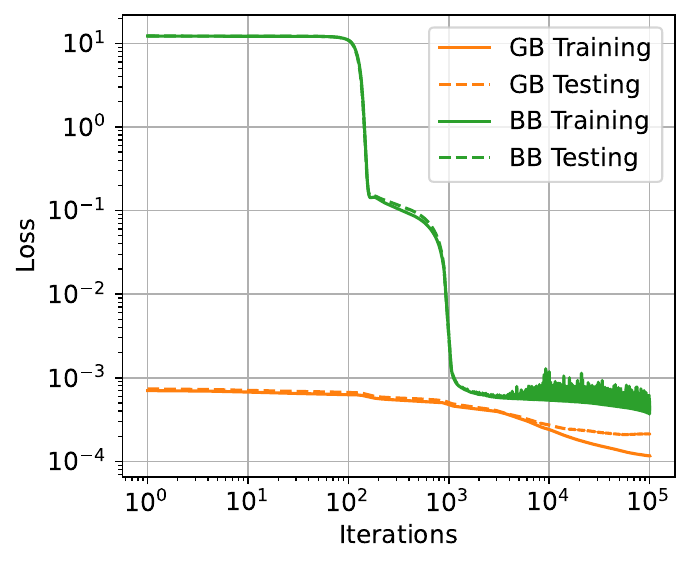}
    \caption{\textbf{The results of model training over the experimental dataset.} The loss, chosen to be the logarithm of the MSE is evaluated for the training and testing examples at each iteration for the GB and BB models.}
    \label{fig:model_training}
\end{figure}
 
\begin{figure*}
    \centering
    \includegraphics[width=\linewidth]{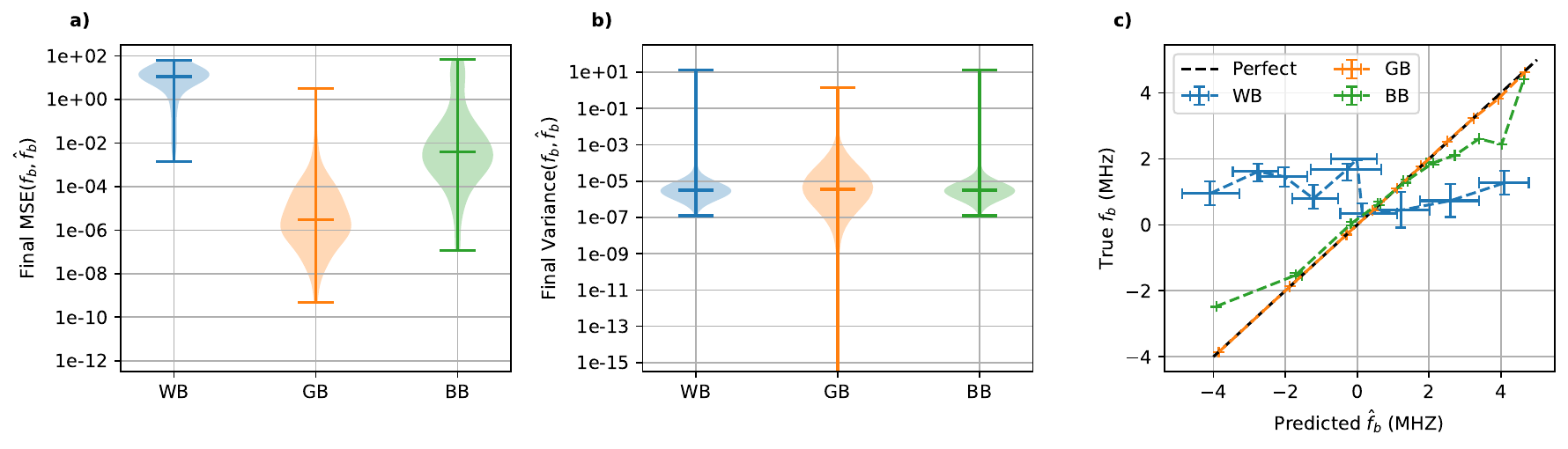}
    \caption{\textbf{The results of estimating unknown Larmor frequencies.} a) A violin plot of the MSE at the final iteration of the Bayesian procedure using the uniform prior, and b) the variance of the estimate,  averaged over randomized ordering of the data for the WB, GB, and BB models. The horizontal lines represent minimum, median, and maximum, while the violin represents a kernel density estimate. c) The calibration plot for each of the three models as compared to perfectly calibrated model (i.e. $\hat{f_b}=f_b$). The error bars represent 95\% confidence over the 100 runs of the Bayes procedure, where each run is a randomized permutations of the measurement sequence.
    }
    \label{fig:stats}
\end{figure*}

\paragraph*{Protocol Implementation:} The proposed approach is implemented in Python, with the WB, GB, and BB implemented using the Tensorflow ~\cite{tensorflow} and Keras ~\cite{keras} packages. The dataset is randomly split into training and testing with ratio 90/10, (i.e. 8688  examples are used for training, and 964 examples are used for testing). The GB and the BB are both trained for $10^{5}$ iterations, and then used within the Bayesian estimation procedure. We have not included any additional experimental parameters, and thus $\chi$ is an empty vector. The learning rate used for the GB is $10^{-5}$ while for the BB is $10^{-4}$. We chose a higher learning rate for the BB to make sure the loss is comparable to that of the GB at the end of the iterations. Figure \ref{fig:model_training} shows the training and testing losses for both architectures versus the iterations. 

The examples in the full dataset, i.e. training and testing examples are then re-grouped by the Larmor frequency $f_B$, to yield 159 sets consisting of the tuples  $(\tau,\phi,\pi_0,\pi_1,P_{\text{cl}})$. The number of tuples in each set vary. The Bayesian procedure is then applied to each set, by passing sequentially the tuples to update the posterior probability of the estimated $\hat{f}_B$. We use $M=5000$ points for discretizing the interval. The true estimate (mean of the posterior distribution), and variance are computed at each iteration (i.e. after one tuple is passed). The MSE between the true and estimate frequencies is also recorded. The whole procedure is then repeated 100 times, where the order of the tuples processed is randomized. The estimate, variance, and MSE is then averaged over this randomized sequences. While Bayesian estimation is order-independent at convergence, the intermediate trajectories differ. This can be used to find error bars on the estimate. To provide a benchmark to which we can compare the performance of the GB, we utilize a commonly-accepted WB model described using Equation \ref{eq:likelihood}, with $T_2^*=5.4 \ \mu$s (as calibrated from the experiment), and the pure BB described earlier. 

To generate a calibration plot, we follow the following procedure. We divide the range of the estimated Larmor frequencies from the dataset $\hat{f}_b$ into 10 bins based on ascending ordering. We also track the ground truth $f_b$ associated to each estimate $\hat{f_b}$. For each bin, we take the average over the estimates, giving the x-coordinate of one point on the calibration plot, while the average over the true frequencies for that bin becomes the y-coordinate. We can then repeat the last step for each bin, and plot all the points relative to the perfect calibration line $f_b=\hat{f_b}$. We do this for the estimates obtained from the GB, as well as the WB and BB. In order to get error bars, we recompute the calibration plot for each of the 100 runs of the Bayesian procedure where the ordering of the data is permuted. The 95\% confidence is then computed for both the x- as well as the y-coordinates of each calibration point. 

\begin{figure*}
    \centering
    \includegraphics[width=\linewidth]{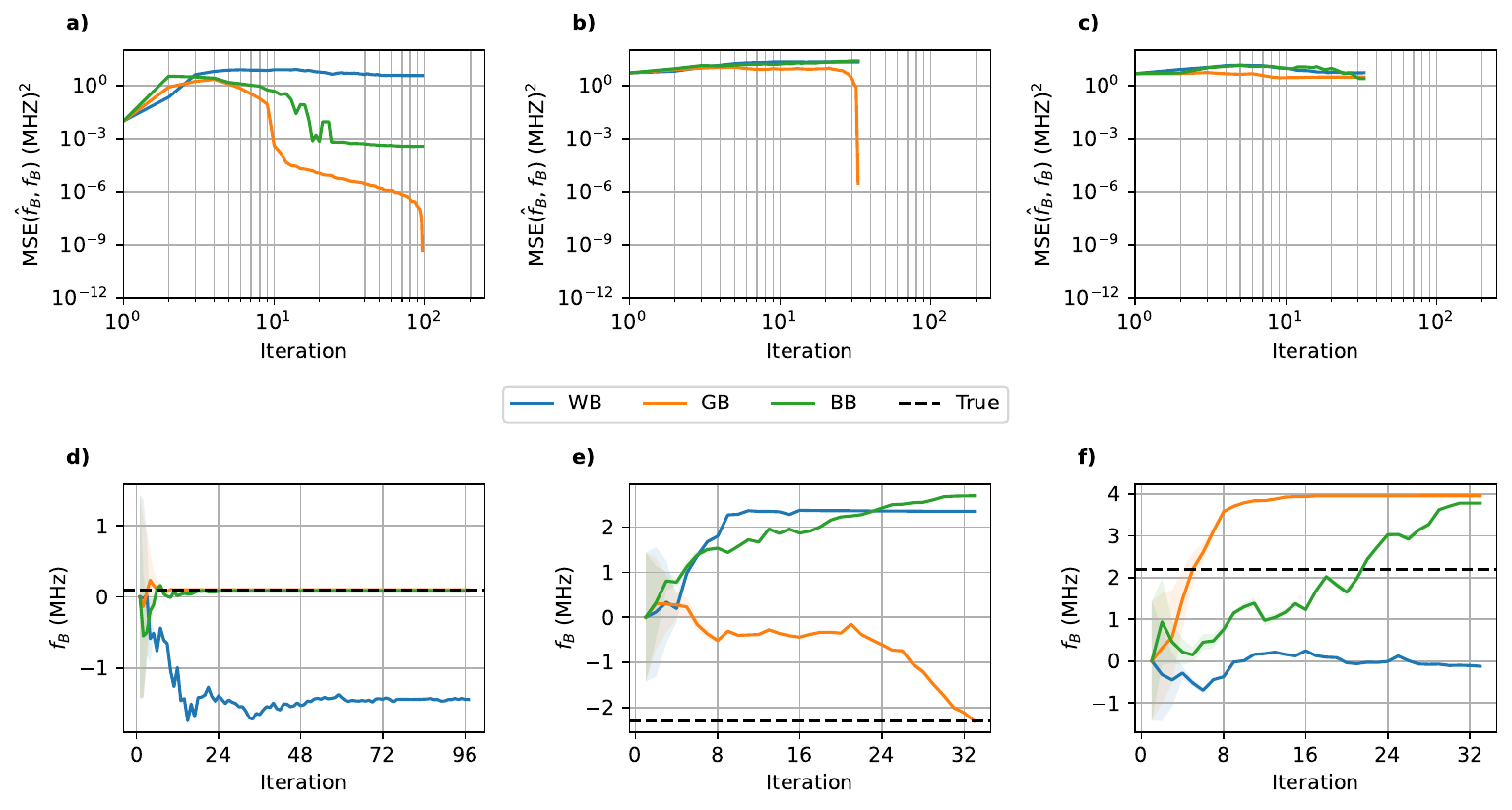}
    \caption{\textbf{Examples of the Bayesian estimation procedure for best-case, average-case, and worst-case}. The MSE of the estimation at each iteration averaged over the randomization of the sequence is plotted versus iteration for a) best-case b) average-case, and c) worst-case.The estimated frequency is then plotted at each iteration for d) best-case e) average-case, and f) worst-case. The dotted line represents the true frequency. The performance is compared for the WB, GB and BB models.}
    \label{fig:Estimation}
\end{figure*}

Figure \ref{fig:stats} summarizes the statistics of the results of running the experiments using the uniform prior, while Supplementary Figure 1 shows a similar plot for the Gaussian prior. In both Figures, we show the distribution of the MSE (averaged over randomization of tuples) over all frequencies, at the final iteration. We also show a similar plot for the variance distribution,  as well as the calibration plot for assessing the reliability of the Bayesian protocol.  Figure \ref{fig:Estimation} shows the performance of the best-case, an average-case, and the worst case of the GB, compared with the WB and BB performance for the uniform prior. The first row in the plot shows the averaged MSE per iteration for each of the three cases. The second row shows the estimate and the true frequency per iteration for the three cases. Supplementary Figure 2 shows a similar plot for the Gaussian prior. Finally, we show in Figure \ref{fig:posterior}, the posterior distribution after the first and the last iteration as compared to the prior, comparing the performance for GB in the best-case against WB and BB. Supplementary Figures 3 and 4 shows similar plots for the average-case, and worst-case examples.

\section{Discussion}
\label{sec:discussion}
In this paper, we proposed a graybox-based approach to improve the performance of NV quantum sensors. The GB allows modeling non-ideal behavior due to state-preparation errors, finite-pulses, and noise. Integrated with a Bayesian procedure, the Larmor frequency of the NV can be estimated.

Figure \ref{fig:model_training} shows a lower training loss for the GB as compared to the BB, at any given iteration. This implies that it is easier and faster to train the GB model to get to a certain performance level compared to a pure BB. For both models, we see that the validation curves are very close to the training curves, indicating that neither model is overfitting. Thus, the size of the neural networks involved in both structures is suitable to the dataset size. 

The experimental results show a significant superior performance of the GB model compared to the WB and BB approaches. In Figure \ref{fig:stats}a, we see that the distribution of the MSE is centered around $10^{-5} \ (\text{MHz})^2$, with few outliers performing worse than $10^{-2} \ (\text{MHz})^2$. On the other hand, the distribution for the WB is centered around $10 \ (\text{MHz})^2$ with the best performing outlier is around $10^{-3} \ (\text{MHz})^2$. The BB performance lies in between, with center lying around $10^{-3} \ (\text{MHz})^{2}$ and best outlier at around $10^{-7} \ (\text{MHz})^{2}$.  The variance plots in Figure \ref{fig:stats}b, show the concentration of the three models around $10^{-6} \ (\text{MHz})^2$, showing that the Bayesian approach does converge for most examples for both models, however the final estimate itself might not be accurate, and hence the importance of studying the MSE as well as in the first plot. A better picture to analyze the performance of the three models is the calibration plot shown in Figure \ref{fig:stats}c. We see that the GB plot is almost identical to the perfect calibration line, with very small error bars. The BB, even though trained on the same dataset with acceptable training performance, had worse performance outside the frequency range -2 MHz to 2 MHz. However, the error bars are still low enough to have confidence about the BB estimates. The WB on the other hand had the worst performance. Even though the predictions span the full range of frequencies in the experiment, there was no match except near 0 MHz. The error bars were also quite high. The superior performance of the GB is also reflected in Figure \ref{fig:Estimation}. We see that in the best-case and average-case, estimate using GB is much closer to the true value of the frequency in comparison to the WB. The worst-case did not converge to the true estimate. This could be attributed to the prediction errors of the GB, due to the limitations on the size of the training dataset. 

We note that convergence of the frequency estimate is governed by two distinct effects. The first is the periodicity of the Ramsey likelihood, which can render the posterior multimodal for broad priors; this is mitigated in our protocol by the geometrically-increasing sensing-time sequence $\tau_k = \tau_0 \cdot 2^k$ (Section IV), which breaks the periodic degeneracy so that the combined likelihood retains a single dominant peak. We did not observe the estimator becoming trapped in spurious local maxima due to this periodicity alone. The dominant source of the convergence failures we did observe is instead model mismatch: when the assumed likelihood (WB, and to a lesser extent BB) does not accurately describe the true experimental $P_{cl}(f_B)$, the posterior can develop a poorly-conditioned or secondary peak, producing the biased, non-converging trajectories seen for the WB model in Fig. ~\ref{fig:Estimation}.

The BB had mixed performance, where it did converge to the true state in the best-case example, but not in the average-case and worst-case. Our experiments did show that moving from 80/20 split to 90/10 split improved the GB performance and consequently the Bayesian estimation. However, upon analyzing the current results, we found that only 3 out of 159 frequencies did not converge using the GB model. These results show the robustness of the GB approach as in modeling the experiment. The inferior performance of the WB model suggests that there are significant effects in the experiments that are not captured by the theory.

\begin{figure*}
    \centering
    \includegraphics[width=\linewidth]{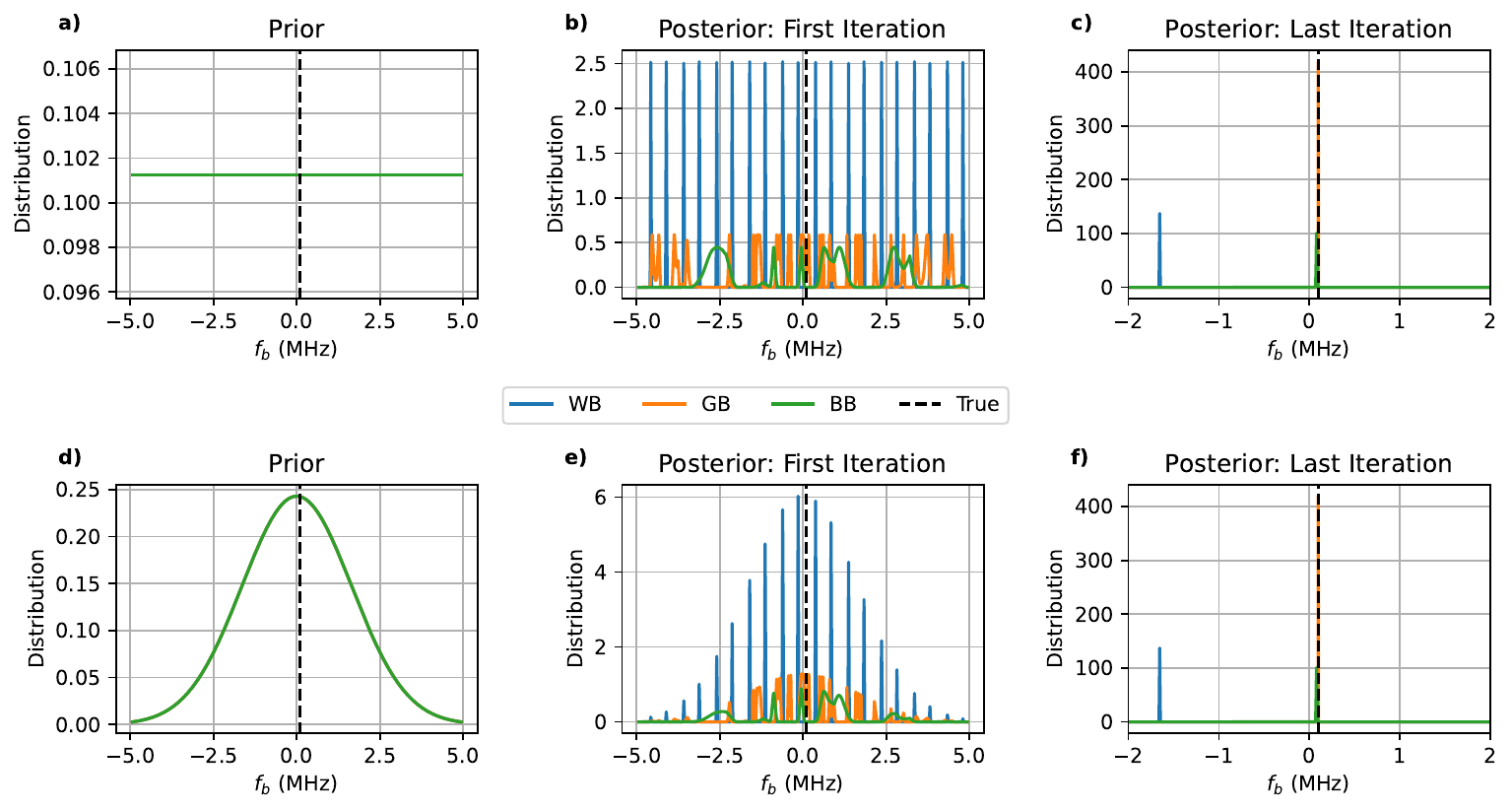}
    \caption{\textbf{The posterior distribution of the estimated Larmor frequency for the best-case example.} The plot is for the first permutation of the measurement sequence, and The performance is compared for the WB, GB, and BB models. The first row is starting from the uniform prior while the second row is starting from Gaussian prior. a) and b) show the prior distribution. c) and d) show the posterior distribution after applying one iteration of the Bayesian procedure. e) and f) show the posterior distribution at the end.}
    \label{fig:posterior}
\end{figure*}

With regards to the effect of choosing the prior distribution, we see that going from uniform distribution to Gaussian distribution did not change the overall results. By comparing Figure \ref{fig:stats} with Supplementary Figure 1, we see the statistics are the same. Looking at Figure \ref{fig:posterior} and Supplementary Figures 3 and 4, we see that the posterior changed after one iteration of the Bayesian update. However, at the end of the iterations, both priors converge to the same posterior. This shows that the prior did not affect the performance of the protocol. Moreover, we see that at the last iteration, the distribution is largely concentrated around some Larmor frequency. This shows that the Bayesian procedure itself is performing as expected, and that the prediction accuracy of the model is the key factor in determining the overall performance of the sensing protocol.

Because inference is performed by direct evaluation of the (unnormalized) posterior over a fine frequency grid, rather than via gradient-based or Laplace-type approximations, the procedure does not require inverting a Hessian matrix and is therefore not susceptible to the ill-conditioning that weak parameter identifiability can cause in those methods. A well-identified frequency corresponds to a sharply peaked posterior (Fig. ~\ref{fig:posterior}), whereas poor identifiability manifests as a multimodal distribution. This  provides, in principle, a diagnostic that could be used to flag unreliable estimates in future adaptive implementations.

One of the critical points in deploying schemes involving deep learning architectures is the datasets required for training, particularly the data acquisition timescales for the training set. While GB approaches provide a significant improvement compared to fully data-driven approaches, they still require a sizable training dataset. For the example discussed here, collecting about 10,000 datapoints took a couple of weeks, taking into account periodic setup re-alignment, re-calibration of the different steps and occasional drifts and equipment failures. 

Enhancing the optical collection efficiency and improving the stability of the experimental setup can substantially accelerate the acquisition of training datasets. For instance, embedding the NV center in photonic structures ~\cite{hadden_StronglyEnhancedPhoton_2010, li_EfficientPhotonCollection_2015, kim_ScalableNanoscalePositioning_2025} can increase the photon count rate to beyond $10^6$~cps, yielding an average of approximately 0.2 detected photons in a 200~ns readout window. Assuming detection probabilities of $\pi_0 = 0.03$ and $\pi_1 = 0.02$ per repetition, 10,000 experimental repetitions produce average photon counts of 300 and 200 for the $m_s = 0$ and $m_s = 1$ states, respectively. These distributions are separated by more than three standard deviations, ensuring that the states are readily distinguishable. Given an average Ramsey experiment duration of 5~$\mu$s, a single experiment with 10,000 repetitions requires only about 0.05~s. Consequently, acquiring a dataset of 10,000 points can be accomplished in under one hour (neglecting the time required to optimize the NV center position, calibrate pulses, etc). This rapid acquisition enables frequent re-calibrations of the GB model for the quantum sensor, thereby maintaining high reliability and consistent performance over time.

There are several routes for future work to deploy this technique into quantum sensing applications. The first point is to extend this work to more complex experiments, for example the long multi-pulse sequences used in nanoscale magnetic resonance ~\cite{budakian_RoadmapNanoscaleMagnetic_2024, vandestolpe_Mapping50spinqubitNetwork_2024, belliardo_MultidimensionalQuantumEstimation_2025}. Such pulse sequences consist of repeated units, for example $\left(  \tau - \pi - \tau \right)^N$ with $N$ as large as 64, 256 or more. Because noise and pulse imperfections are expected to be consistent across all units, one would train a smaller BB block tailored to the base unit rather than a single, large model for the entire long sequence. Therefore, we expect the computational requirements for this important application, both in terms of training-set size and training time, to be comparable to those reported here. 

A second avenue for future research is the integration of GB models with real-time Bayesian experiment design~\cite{rainforth_ModernBayesianExperimental_2024}. In this configuration, the algorithm decides the next optimal settings to maximize the information extracted from each measurement, using the current probability distribution for the parameter of interest. This is heavily dependent on the model used for the system, and a GB model which represents the real system more faithfully, as shown here, is expected to be advantageous. As the GB model is fully differentiable, it is possible to deploy online experiment design by maximizing information-theoretic quantities such as Fisher information~\cite{scerri_ExtendingQubitCoherence_2020a, craigie_ResourceefficientAdaptiveBayesian_2021, zohar_RealtimeFrequencyEstimation_2023a,arshad_RealtimeAdaptiveEstimation_2024b}, or deploying more sophisticated procedures such as model-aware reinforcement learning~\cite{belliardo_model-aware_2024, belliardo_applications_2024}. From an experimental point of view, this requires the evaluation of the trained GB model on timescales much shorter than the measurement time: this should be feasible with modern 
arbitrary waveform generators integrating GPU hardware.

From a broader perspective, this work could be further extended by integrating physically interpretable, data‑driven frameworks for open quantum systems, such as control-metric approaches to non-Markovian dynamics~\cite{mayevsky_ResourceboundedControllabilityBenchmarking_2026} and quantum feature-space representations of qubit–bath interactions~\cite{wise_QuantumFeatureSpace_2025}. Overall, this study provides, to the best of our knowledge, the first experimental demonstration of a gray‑box modeling approach applied to a solid‑state open quantum system, laying the groundwork for scalable and interpretable methodologies in quantum sensing.

A related extension to this work, is the real-time estimation of the noise operator through the use of online estimation techniques. A different approach such as~\cite{wise_QuantumFeatureSpace_2025} could be more suitable in that case. This could be beneficial for tracking long-term drifts in system parameters, a current challenge in all experimental platforms. 

A different possibility to adapt the trained model for small time-dependent drifts in the noise and imperfections is to utilize transfer learning methods ~\cite{pan_SurveyTransferLearning_2010, yosinski_HowTransferableAre_2014, wang_TransferLearningPhysicsInformed_2025, kirchmeyer_GeneralizingNewPhysical_2022}. Such techniques preserve the mechanistic structure of the model, while enabling fast adaptation to slightly different noise and imperfection, with limited training data. These techniques could also be used to adapt the quantum sensor to different configurations or environments, for example using a multi-head network where a shared core is combined with multiple environment- or system-specific heads ~\cite{chen_HardwareConditionedPolicies_2018, pellegrin_TransferLearningPhysicsInformed_2022, wu_PhysicsInformedMultitaskLearning_2023}, or conditioning the dynamical models on contextual parameters, specific to each environment, e.g. by a hyper-network ~\cite{ha_HyperNetworks_2016} learned jointly with a context vector from observed data ~\cite{kirchmeyer_GeneralizingNewPhysical_2022}.\\

\paragraph*{Author Contributions:}
A.Y. designed, implemented, and trained the graybox; implemented the Bayesian procedure, and generated the results. S.T., P.M., M.J.A., and N.W. conducted all the experimental work on the NV setup and collected the dataset. A.P. and C.B. supervised the work. All authors contributed in writing the paper. \\ 

\paragraph*{Data and Code Availability:} The data and code used in this work are publicly available at \url{https://github.com/akramyoussry/GB-QSensing}. \\

\paragraph*{Acknowledgments:}
A.P. and C.B. equally led this work. We thank Ben Haylock and Christiaan Bekker for useful discussions. We acknowledge funding by the Engineering and Physical Sciences Research Council (EP/S000550/1, EP/V053779/1,  EP/T00097X/1, EP/Z533208/1, EP/Z533191/1, UKRI2706), the European Innovation Council (QuSPARC, grant agreement 101186889) and the Australian Government through the Australian Research Council under the Centre of Excellence scheme (No: CE170100012). This work is also supported by the project 23NRM04 NoQTeS, which has received funding from the European Partnership on Metrology, co-financed from the European Union's Horizon Europe Research and Innovation Programme and by the Participating States, and has benefited from resources from the National Computational Infrastructure (NCI Australia), an NCRIS enabled capability supported by the Australian Government. AP acknowledges an RMIT University Vice-Chancellor's Senior Research Fellowship and a Google Faculty Research Award.
\bibliography{references, references_zotero}
\end{document}


\title{Supplementary Materials for: Bayesian quantum sensing using graybox machine learning}

\author{Akram Youssry}
\address{Quantum Photonics Laboratory and Centre for Quantum Computation and Communication Technology, RMIT University, Melbourne, VIC 3000, Australia}
\affiliation{School of Electrical Engineering and Telecommunications, UNSW Sydney, Sydney, NSW 2052, Australia}

\author{Stefan Todd}
\address{
Institute of Photonics and Quantum Sciences, SUPA,
School of Engineering and Physical Sciences, Heriot-Watt University, Edinburgh EH14 4AS, UK}

\author{Patrick Murton}
\address{
Institute of Photonics and Quantum Sciences, SUPA,
School of Engineering and Physical Sciences, Heriot-Watt University, Edinburgh EH14 4AS, UK}

\author{Muhammad Junaid Arshad}
\address{
Institute of Photonics and Quantum Sciences, SUPA,
School of Engineering and Physical Sciences, Heriot-Watt University, Edinburgh EH14 4AS, UK}

\author{Nicholas Werren}
\address{
Institute of Photonics and Quantum Sciences, SUPA,
School of Engineering and Physical Sciences, Heriot-Watt University, Edinburgh EH14 4AS, UK}

\author{Alberto Peruzzo}
\email{alberto.peruzzo@gmail.com}
\address{Quantum Photonics Laboratory and Centre for Quantum Computation and Communication Technology, RMIT University, Melbourne, VIC 3000, Australia}
\address{Quandela, Massy, France}

\author{Cristian Bonato}
\email{c.bonato@hw.ac.uk}
\address{
Institute of Photonics and Quantum Sciences, SUPA, School of Engineering and Physical Sciences, Heriot-Watt University, Edinburgh EH14 4AS, UK}

\maketitle

\section{Supplementary Note 1: State Preparation Error Modeling}

In this section, we show that the stochastic Hamiltonian $H_\text{Prep}(t)=\beta(t)\sigma_x$, where $\beta(t)$ is a white Gaussian stationary random process with zero mean and variance (power) $A_p$ results in the non-unitary evolution described by a bit-flip channel $\mathcal{E}(\ket{0}\!\!\bra{0})=(1-\epsilon)\ket{0}\!\!\bra{0} + \epsilon \ket{1}\!\!\bra{1}$. 

The resulting unitary is given by
\begin{align}
    U_{\text{prep}}(t)&=\mathcal{T}e^{-i\int_0^tH_{\text{prep}}(s)ds}\\
    &=e^{-i\int_0^tH_{\text{prep}}(s)ds} \\
    &=e^{-i\int_0^t\beta(s)ds}\ket{+}\!\!\bra{+} + e^{i\int_0^t\beta(s)ds}\ket{-}\!\!\bra{-}
\end{align}
In the second line, we notice that $[H_{\text{prep}}(t_1),H_{\text{prep}}(t_2)]=0, \ \forall t_1,t_2$, and with this the time-ordered operator $\mathcal{T}$ can be removed. In the third line we applied the eigendecomposition of $\sigma_x$ Pauli matrix, and applied the diagonalization to calculate the matrix exponential. Now, starting from the initial state $\rho(0)=\ket{0}\!\!\bra{0}$, the evolution under the Hamiltonian is given by
\begin{align}
    \rho(T_p)&=\left\langle U_{\text{prep}}(T_p)\rho(0)U^{\dagger}_{\text{prep}}(T_p)\right \rangle_{\beta} \\
    &=\left\langle\left(e^{-i\int_0^{T_p}\beta(s)ds}\ket{+}\!\!\bra{+} + e^{i\int_0^{T_p}\beta(s)ds}\ket{-}\!\!\bra{-} \right) \ket{0}\!\!\bra{0}\left(e^{i\int_0^{T_p}\beta(s)ds}\ket{+}\!\!\bra{+} + e^{-i\int_0^{T_p}\beta(s)ds}\ket{-}\!\!\bra{-} \right) \right\rangle_{\beta} \\
    &= \frac{1}{2}\left\langle \ket{+}\!\!\bra{+} + e^{-2 i\int_0^{T_p}\beta(s)ds} \ket{+}\!\!\bra{-} +  e^{2 i\int_0^{T_p}\beta(s)ds}\ket{-}\!\!\bra{+} + \ket{-}\!\!\bra{-} \right \rangle_{\beta} \\
    &=\frac{1}{2}\left(\mathbb{I} + \Gamma \ket{+}\!\!\bra{-} + \Gamma^* \ket{-}\!\!\bra{+}\right) \\
    &= \begin{pmatrix}
        1 + \Gamma + \Gamma^* &  -\Gamma +\Gamma^*\\ \Gamma -\Gamma^* & 1 - \Gamma - \Gamma^*
    \end{pmatrix}.
\end{align}
Here we defined 
\begin{align}
    \Gamma :&= \left\langle e^{-2 i\int_0^{T_p}\beta(s)ds} \right\rangle_{\beta} \\
    &=e^{-2\int_0^{T_p}\int_0^{T_p}\braket{\beta(t_1}\beta(t_2)dt_1dt_2  } \\
    &= e^{-2\int_0^{T_p}\int_0^{T_p}R_{\beta}(t_1-t_2) dt_1dt_2  } \\
    &= e^{-2\int_{-\infty}^{\infty}\int_{-\infty}^{\infty}A_p\delta(t_1-t_2)y(t_1)y(t_2) dt_1dt_2  } \\
    &=e^{-2A_p\int_{-\infty}^{\infty}y(t_2)^2dt_2  }\\ 
    &= e^{-2 A_P T_p  }
\end{align}
In the second line, we applied the cumulant expansion of the process $\int_0^{T_p}\beta(s)ds$, using the fact that $\beta(t)$ is a zero-mean Gaussian random process. In the third line, we used the stationarity of  $\beta(t)$, thus the correlation function $R_\beta(t_1,t_2)$ only depends on the difference between $t_1$ and $t_2$. In the fourth line, we used the fact that $\beta(t)$ is white noise (uncorrelated) and introduced the window function $y(t)$ defined as
\begin{align}
    y(t)=\begin{cases}
        1, \quad t \in [0,T_p]\\
        0, \quad \text{otherwise}.
    \end{cases}
\end{align}
Since $\Gamma\in\mathbb{R}$, we can reduce the expression for the evolved state to
\begin{align}
     \rho(T_p)&=\frac{1}{2}\begin{pmatrix}
        1 + 2\Gamma &  0\\ 0 & 1 - 2\Gamma
    \end{pmatrix}\\
    &=(1-\epsilon)\ket{0}\!\!\bra{0} + \epsilon\ket{1}\!\!\bra{1},
\end{align}
if we choose 
\begin{align}
    \Gamma = \frac{1}{2} -\epsilon \implies A_pT_p=-\frac{1}{2}\log\left(\frac{1}{2}-\epsilon\right). 
\end{align}
Since the target is to find an effective equivalent model for the state preparation error, we can choose the noise power $A_P$ and evolution time $T_p$ as needed to yield the desired mixed state $\mathcal{E}(\ket{0}\!\!\bra{0})$.  \qed

\newpage
\section{Supplementary Figures}

\begin{figure}[H]
    \centering
    \includegraphics[width=0.85\linewidth]{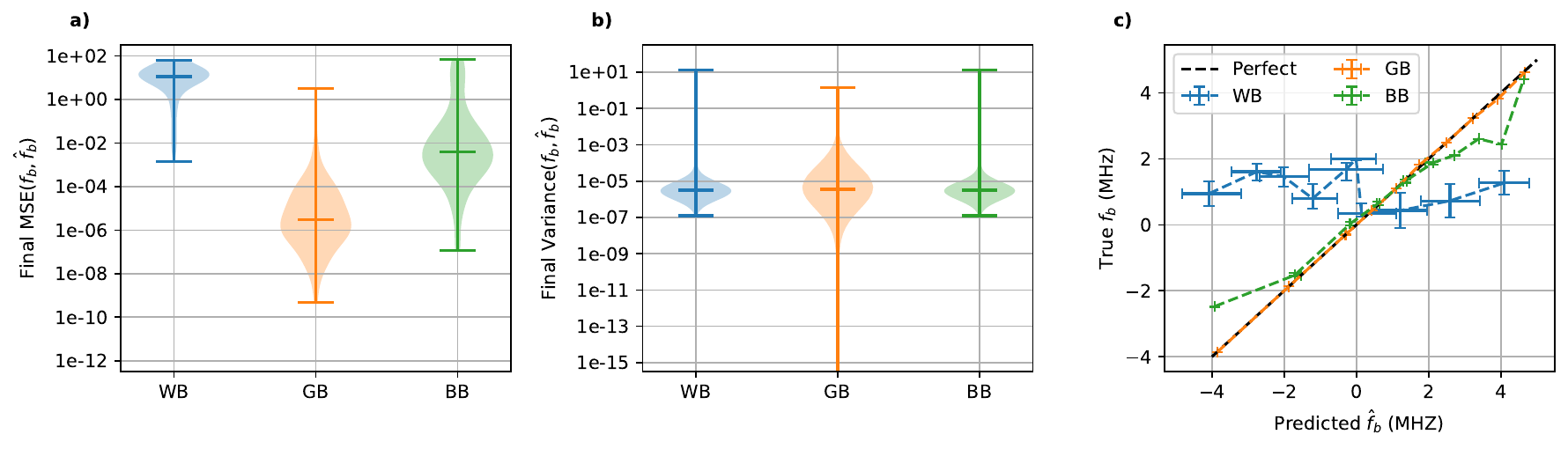}
    \caption{\textbf{The results of estimating unknown Larmor frequencies.} a) A violin plot of the MSE at the final iteration of the Bayesian procedure using the Gaussian prior, and b) the variance of the estimate,  averaged over randomized ordering of the data for the WB, GB, and BB models. The horizontal lines represent minimum, median, and maximum, while the blob represent a kernel distribution estimation of the data. c) The calibration plot for each of the three models as compared to perfect calibrated model (i.e. $\hat{f_b}=f_b$).  The error bars represent 95\% confidence over the 100 runs of the Bayes procedure, where each run is a randomized permutations of the measurement sequence. }
\end{figure}

\begin{figure}[H]
    \centering
    \includegraphics[width=\linewidth]{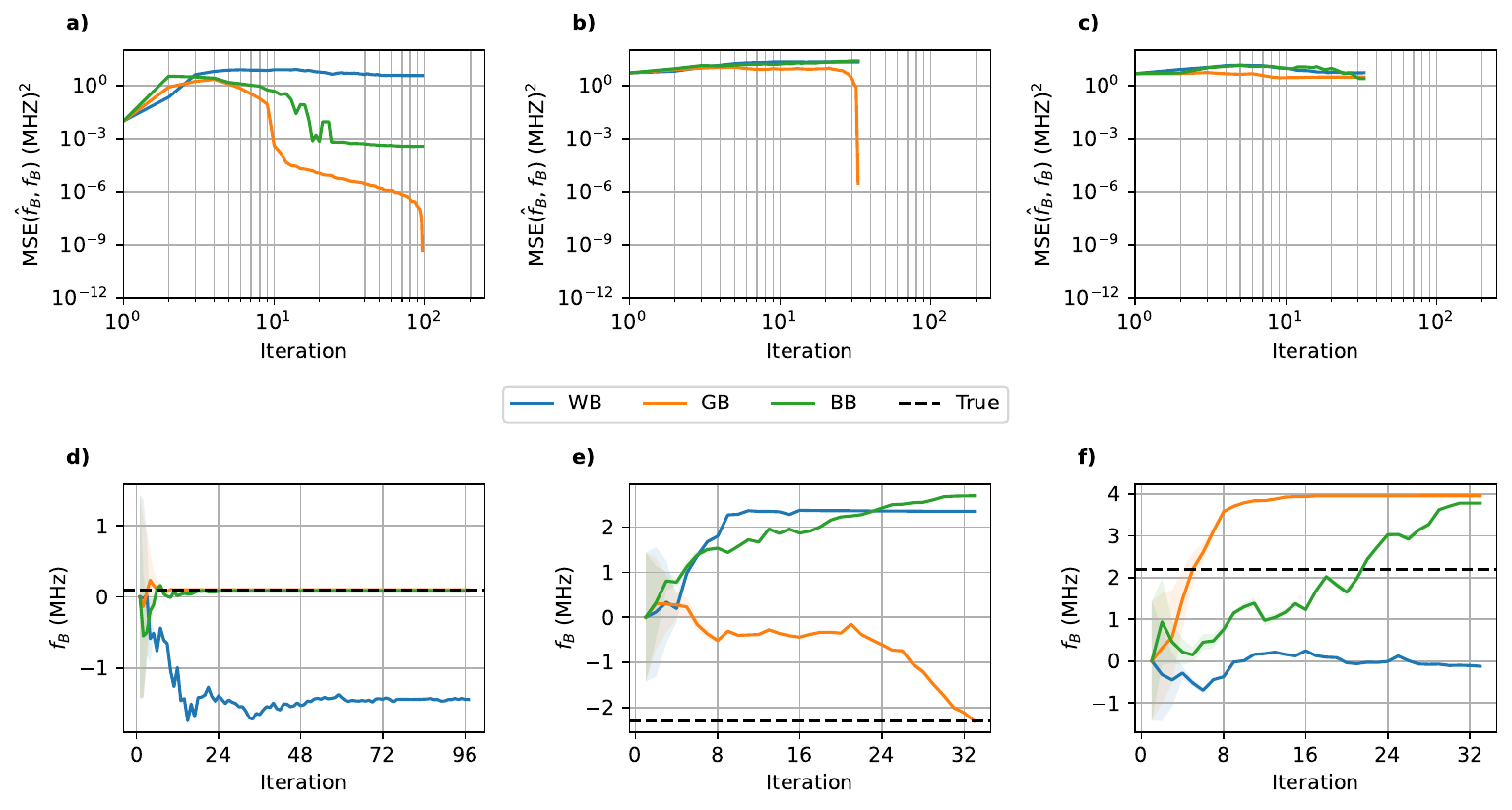}
    \caption{\textbf{Examples of the Bayesian estimation procedure for best-case, average-case, and worst-case}. The MSE of the estimation at each iteration averaged over the randomization of the sequence is plotted versus iteration for a) best-case b) average-case, and c) worst-case.The estimated frequency is then plotted at each iteration for d) best-case e) average-case, and f) worst-case. The dotted line represents the true frequency. The performance is compared for the WB,  GB, and BB models.}
\end{figure}

\begin{figure}[H]
    \centering
    \includegraphics[width=\linewidth]{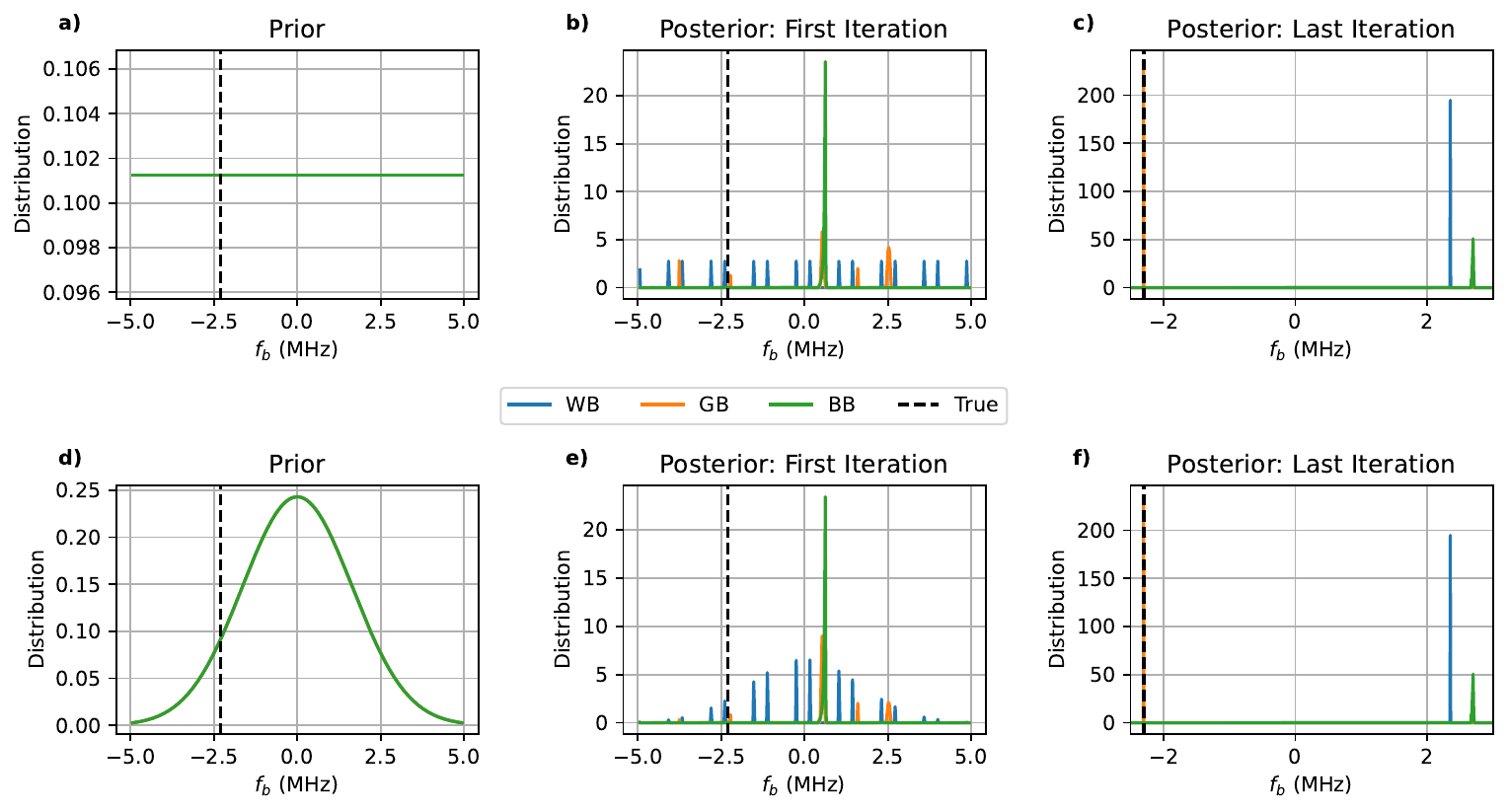}
    \caption{\textbf{The posterior distribution of the Larmor frequency estimated Larmor frequency for the average-case example.} The plot is for the first permutation of the measurement sequence, and the performance is compared for the WB, GB, and BB models. The first row is starting from the uniform prior while the second row is starting from Gaussian prior. a) and b) show the prior distribution. c) and d) show the posterior distribution after applying one iteration of he Bayesian procedure. e) and f) show the posterior distribution at the end.}
\end{figure}

\begin{figure}[H]
    \centering
    \includegraphics[width=\linewidth]{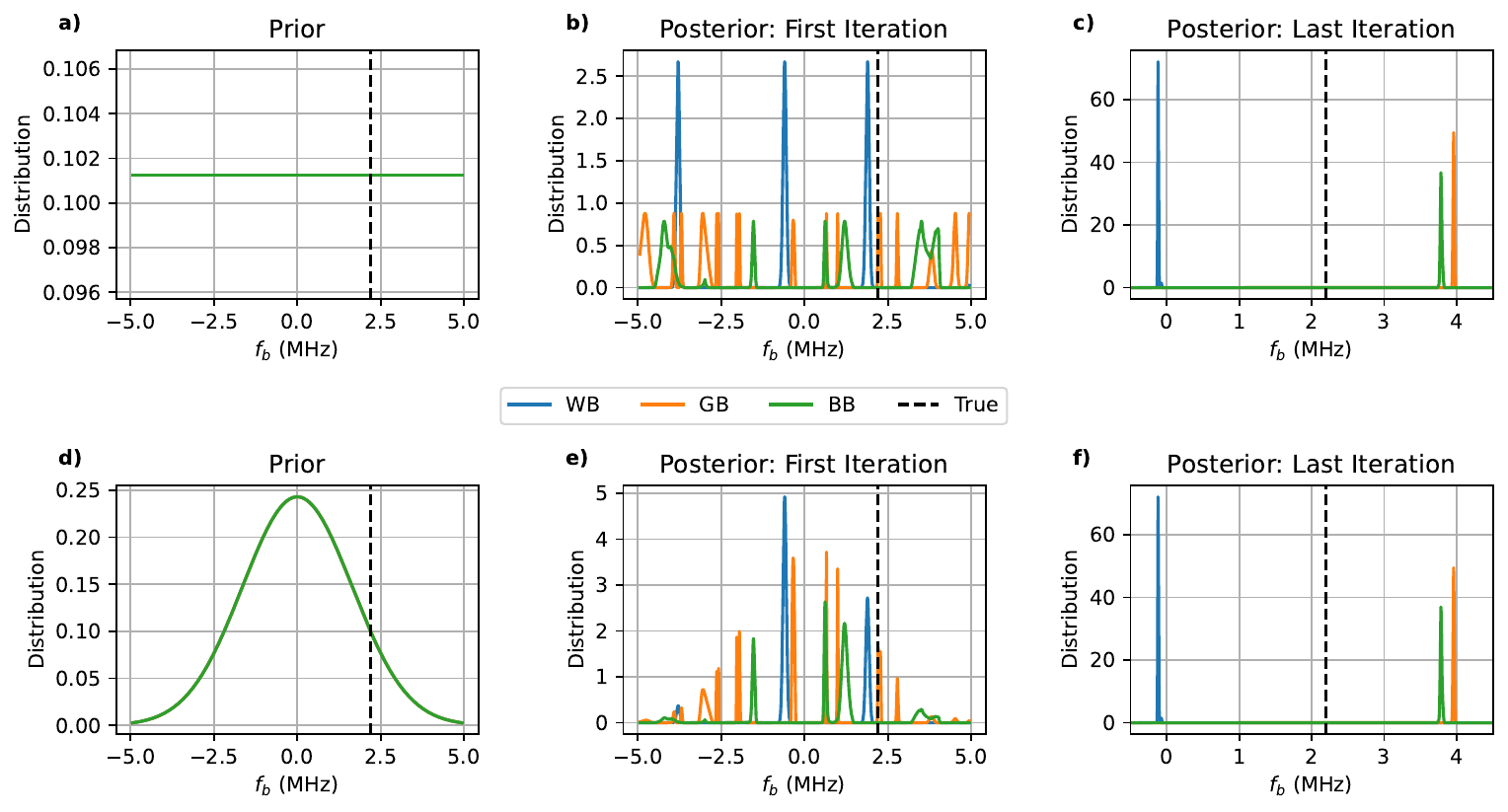}
    \caption{\textbf{The posterior distribution of the Larmor frequency estimated Larmor frequency for the worst-case example.} The plot is for the first permutation of the measurement sequence, and the performance is compared for the WB, GB, and BB models. The first row is starting from the uniform prior while the second row is starting from Gaussian prior. a) and b) show the prior distribution. c) and d) show the posterior distribution after applying one iteration of he Bayesian procedure. e) and f) show the posterior distribution at the end.}
\end{figure}